\documentstyle[aps,preprint]{revtex}

\begin{document}
\draft
\preprint{US-FT/4-96}

\title{Searching for non-minimally coupled scalar hairs}

\author {Alberto Saa\cite{cnpq}}

\address{Departamento de F\'\i sica de Part\'\i culas\\
       Universidade de Santiago de Compostela\\
       15706 Santiago de Compostela, Spain}
\date{\today}
\maketitle

\begin{abstract}
In this work we study  the asymptotically flat, static, and
spherically symmetric black-hole 
solutions of the theory described by the action 
$$S = \int d^nx\sqrt{-g} \left\{\left(1-\xi\phi^2 \right)R 
- g^{\mu\nu}\partial_\mu\phi\partial_\nu\phi\right\},$$
with $n>3$ and arbitrary $\xi$. We demonstrate the absence of
scalar hairs for $\xi<0$.  For $\xi>\xi_c=\frac{n-2}{4(n-1)}$,
we show that there is no scalar hair obeying
$|\phi(r)| < 1/\sqrt{\xi}$ or $|\phi(r)| > 1/\sqrt{\xi}$.
For $0<\xi<\xi_c$, we prove the absence of scalar hairs such that
$|\phi(r)| < 1/\sqrt{\xi}$ or 
$\frac{1}{\xi} < \phi^2(r) < \frac{\xi_c}{\xi(\xi_c-\xi)}$. 
\end{abstract}
\pacs{97.60.Lf, 04.70.Bw}

The Bekenstein black-hole solution\cite{b} for Einstein
 gravity conformally coupled
to a scalar field in four dimensions has had a prominent role in gravitational
Physics. It has an extremal Reissner-Nordstr\"om  geometry and an
$\frac{1}{r}$-type scalar field, and it was one of the first counter-examples
to the ``no-hair'' conjecture\cite{RW}. The scalar field diverges in the
horizon, and such a divergence is crucial to the violation of the
no-hair theorems, as recent works have revealed\cite{Z,B1,Saa}.
The Bekenstein solution is
an asymptotically flat, static, and
spherically symmetric solution for the theory described by
the action
\begin{equation}
\label{action}
S[g,\phi] = \int d^nx\sqrt{-g} \left\{\left(1-\xi\phi^2 \right)R
- g^{\mu\nu}\partial_\mu\phi\partial_\nu\phi\right\}
\end{equation} 
with $\xi=\frac{1}{6}$ and $n=4$. 
The coupling defined by such values is
called conformal because with them the action
 (\ref{action}) is invariant under the
map defined by the conformal transformation 
$g_{\mu\nu} = \Omega^2\bar{g}_{\mu\nu}$, $\Omega^2>0 $, 
and by the field transformation
$\phi = \Omega^{-1}\bar{\phi}$. This map
 can be easily extended for $n$-dimensional space-times;
with the coupling given by $\xi_c = \frac{n-2}{4(n-1)}$, the action
(\ref{action}) is
conformal invariant, with the field redefinition given by
$\phi = \Omega^{\frac{2-d}{2}}\bar{\phi}$. 

Although we know that scalar fields are not elementary fields
in nature, they commonly arise in effective actions. 
In fact, some scalar actions have been considered recently in 
astrophysical contexts, see for instance \cite{B3}.
However,
with the conformally coupled case as the only exception\cite{Z,XZ,xd,CK}, 
only minimally
coupled scalar fields have been examined. 
In \cite{B1} it is presented a new theorem which rules out a multicomponent
scalar hair with non-quadratic Lagrangian, but with minimal coupling to 
gravity.
As it is stressed in 
 \cite{B1}, scalar fields effective actions are obtained by integrating
the functional integral of the elementary fields in nature over some of
the fields, and more complicated actions involving non-minimally coupling
should arise.

In this line, the first natural question that can arise in the analysis of
(\ref{action}) is if there exists another non-vacuum ($\phi$ not constant) 
black-hole 
for some $\xi$ or $n$. Such an investigation has already been started,
and the up-to-now data is the following. Long standing results
state that for the minimally coupled
case, $\xi=0$, there is no other black-hole solution than the
vacuum Schwarzschild one for $n>3$\cite{S,V,Sp,X1}. For the conformally
coupled case, $\xi=\xi_c$, recent works have established that the
only static and spherically symmetric 
non-vacuum black-hole solution of (\ref{action}) with $n>3$
is the four dimensional Bekenstein
one\cite{XZ,xd,CK}. 
The present work points toward the conclusion of this investigation
by showing that the only black-hole solutions of (\ref{action})
are the Schwarzschild ones for $n>3$ and for
very large ranges of $\xi$ and $\phi$.
Hereafter, we 
will use the term black-hole solution to denote  
an asymptotically flat, static, and
spherically symmetric black-hole solution. 
We show that the Schwarzschild black-hole solution is the only one 
for $\xi<0$. For $\xi>\xi_c$, we prove the
absence of scalar hairs obeying
$|\phi(r)| < 1/\sqrt{\xi}$ or $|\phi(r)| > 1/\sqrt{\xi}$.
We also demonstrate the absence of scalar hairs obeying
$\frac{1}{\xi} < \phi^2(r) < \frac{\xi_c}{\xi(\xi_c-\xi)}$
or $|\phi(r)| < 1/\sqrt{\xi}$
for $0<\xi<\xi_c$. 
These results are in agreement with the recent results about
the uniqueness of the four dimensional Bekenstein black-hole 
solution\cite{Z,XZ,xd,CK}.

The demonstration of our results centers in a covariant method for
generating solutions for (\ref{action}) starting from the well
known solutions of the minimally coupled case.  It
generalizes the method for generating solutions for the conformally
coupled case in $n>3$ dimensions presented in \cite{xd}. For
$n=4$, our method reproduces the method used in \cite{AWKP} for generating
solutions for arbitrary $\xi$ in four dimensions. Such methods are
based in conformal transformations and $\phi$-redefinitions
and they have a long history. The Ref. \cite{MS}, for instance,
 presents a good set of references on the subject. A
method of this type was early presented by Bekenstein\cite{b} and used by him
for generating solutions for the conformally coupled case starting from
the minimally coupled one in four dimensions; this was the way that
he  obtained his black-hole solution with conformal hair. We notice
also that Maeda in \cite{m} has used very close machinery to show
that the
action given by $\int d^4x\sqrt{-g}\left\{ 
F(\phi,R) - g^{\mu\nu}\partial_\mu\phi\partial_\nu\phi\right\}$ is
equivalent to an Einstein-Hilbert action plus minimally coupled
self-interacting scalar
fields, equivalent in the sense that there is a conformal transformation
and $\phi$-redefinition connecting them.

The method will provide us with a general asymptotically flat, static,
and spherically symmetric solution of (\ref{action}). Such a general
solution will be given by a two-parameters $(\lambda,r_0)$ family
of solutions, and we will systematically search for values of
$(\lambda,r_0)$ such that the solution corresponds to a black-hole.
To this end, one needs to be capable to identify a black-hole
solution. Due to our experience with the Bekenstein solution
we will not pay attention to possible divergences of the scalar
field $\phi$. We recall that the general static and spherically symmetric
$n$-dimensional metric has the following form in isotropic coordinates
\begin{equation}
\label{metr}
ds^2 = -e^fdt^2 + e^{-h}dr^2 + e^{-h}r^2 d\Omega^2,
\end{equation} 
where $d\Omega^2$ denotes  for the metric of the unitary $(n-2)$ 
sphere.  The metric
(\ref{metr}) describes a black-hole if it has a regular event
horizon, say the hyper-surface $r=r_0$. The necessary and
sufficient conditions
in order to this hyper-surface be a regular horizon are:
\renewcommand{\theenumi}{\roman{enumi}}
\begin{enumerate}
\item $e^f$ vanishes at $r=r_0$, so that the hyper-surface is of the
null type,
\item the invariants of the metric are finite in $r=r_0$.
\end{enumerate}
If condition ii is not verified the hyper-surface $r=r_0$ is said to be
a naked singularity. The invariant of the metric that we will use in
our case is the scalar of curvature $R$, which can 
be obtained from the Euler-Lagrange equations of (\ref{action}), 
\begin{equation}
\label{scal}
R = \frac{1-\xi/\xi_c}{1-\xi(1-\xi/\xi_c)\phi^2}g^{rr} 
\left(\frac{d\phi}{dr}\right)^2.
\end{equation}
We will see that for all candidate solutions to be black-hole $R$
will be singular for $r=r_0$ or such a hypersurface will be not of
the null type, with the only exception of the Schwarzschild solution.

To present the method for generate solutions let us suppose first that
$1-\xi\phi^2(r) > 0$. We can check that
the following transformations
\begin{eqnarray}
\label{trans1}
g_{\mu\nu} &=& (1-\xi\phi^2 )^{-\frac{2}{n-2}} \bar{g}_{\mu\nu}, \\
\label{trans2}
\bar{\phi}(\phi) &=& \int_a^\phi d\chi 
\frac{\sqrt{1 + \xi\left(\frac{\xi}{\xi_c} - 1\right)\chi^2}}{1 -\xi\chi^2},
\end{eqnarray}
on the action (\ref{action}) leads to 
$S[\Omega^2\bar{g},\phi(\bar{\phi})] = 
\bar{S}[\bar{g},\bar{\phi}]$, where 
\begin{equation}
\label{action2}
\bar{S}[\bar{g},\bar{\phi}] = \int d^4x\sqrt{-\bar{g}} 
\left\{\bar{R} - \bar{g}^{\mu\nu}
\partial_\mu\bar{\phi}\partial_\nu\bar{\phi}\right\}
\end{equation}
is the minimally coupled action. Due to the assumption $1-\xi\phi^2(r) > 0$,
the right-handed side of (\ref{trans2}) is a monotonically increasing function
of $\phi$, what guarantees the existence of the inverse $\phi(\bar{\phi})$.
The constant $a$ is to be determined by the boundary conditions of
$\phi$ and $\bar{\phi}$. The conformal transformation (\ref{trans1}) is
valid in general only locally. It is the spherical symmetric in our case
that guarantees that an unique conformal transformation can be used for the
whole black-hole exterior.

The transformation  given by equation (\ref{trans1}) and (\ref{trans2}), 
therefore, maps a solution $(g_{\mu\nu},\phi)$ of (\ref{action}) 
to a solution $(\bar{g}_{\mu\nu},\bar{\phi})$ of (\ref{action2}). The
transformation is independent of any assumption of symmetries, and
in this sense is covariant. Also, we can easily infer that the transformation
is one-to-one in general,
 in the sense that any solution of (\ref{action}) is mapped
in an unique solution of (\ref{action2}). The transformation
preserves symmetries, what means that if
$\bar{g}_{\mu\nu}$ admits a Killing vector $\xi$ such that
${\pounds}_{\xi}\bar{\phi} = 0$,
 then $\xi$ is also a Killing vector
of ${g}_{\mu\nu}$ and ${\pounds}_{\xi}{\phi} = 0$. From this, one concludes
if we know all solutions $(\bar{g}_{\mu\nu},\bar{\phi})$ with a given
symmetry we automatically know all $({g}_{\mu\nu},{\phi})$ with the
same symmetry and obeying $1-\xi\phi^2(r) > 0$.
 We will obtain the general asymptotically flat, static, 
and spherically symmetric solution of (\ref{action}) in this way.

The general asymptotically flat, static, and spherically symmetric
 solution for $n>3$ dimensions $(\bar{g}_{\mu\nu},\bar{\phi})$ 
of the minimally coupled case (\ref{action2}) was derived in \cite{X1}. 
It is given in isotropic coordinates 
by the two-parameter $(\lambda,r_0)$ family of solutions
\begin{eqnarray}
\label{solu}
\bar{\phi} &=& -\sqrt{\frac{n-2}{n-3}(1-\lambda^2)}\ln {\cal R}_n, \nonumber \\
ds^2 &=& \bar{g}_{\mu\nu} dx^\mu dx^\nu =
-{\cal R}^{2\lambda}_ndt^2 + \left( 
1-\frac{r_0^{2n-6}}{r^{2n-6}}
\right)^\frac{2}{n-3} 
{\cal R}^{-\frac{2\lambda}{n-3}}_n 
\left(dr^2 + r^2 d\Omega^2 \right),
\end{eqnarray}
where ${\cal R}_n=\frac{r^{n-3}-r^{n-3}_0}{r^{n-3}+r^{n-3}_0}$.  The parameter
$\lambda$ can take values in $[-1,1]$ in principle, although 
the negative range corresponds to solutions with negative ADM mass\cite{X1}.
For $\lambda = 1$, the solution is the usual exterior $n$-dimensional 
vacuum Schwarzschild 
solution with the horizon at $r'_0=4r_0$, as one can check by using the 
coordinate transformation $r' = r\left(1+\frac{r_0}{r}\right)^2$.
For $0 \le \lambda < 1$, (\ref{solu}) does not represent
a black-hole because 
the surface $r=r_0$ is not a horizon, {\em i.e.} a regular null surface,
but it is instead
a naked singularity, as we can check 
by calculating the scalar of curvature 
\begin{equation}
\label{R}
\bar{R}=
\frac{4(n-2)(n-3)r^{2n-8}r_0^{2n-6}}
     {(r^{n-3}+r_0^{n-3})^{\frac{2(n-2+\lambda)}{n-3}}}
\times \frac{1-\lambda^2}{(r^{n-3}-r_0^{n-3})^{\frac{2(n-2-\lambda)}{n-3}}}. 
\end{equation}

Note that the solution (\ref{solu}) for general $\lambda$
describes only the exterior region ($r>r_0$) of the space-time. 
However, it is still
valid for the interior region for some values of $\lambda$ and $n$. 
In these cases,
the solution for the scalar field is 
$\bar{\phi} = -\sqrt{\frac{n-2}{n-3}(1-\lambda^2)}\ln |{\cal R}_n|$, and
the signature for the interior region could eventually change to
$(+,-,\cdots,-)$. Note that cases like 
$\lambda=0$ and $n=5$ do not correspond to acceptable interior
solutions because the signature in the interior region would be
$(-,-,\cdots,-)$.

For the range $\xi < 0$ the assumption of $1-\xi\phi^2(r)> 0$ is automatically
verified and we can use the transformation (\ref{trans1}) and (\ref{trans2})
for generating the solutions $(g_{\mu\nu},\phi)$ starting from (\ref{solu}).
{}From the regularity of the integrand we have for this case 
$\lim_{\bar{\phi}\rightarrow\infty} \phi(\bar{\phi}) = \infty$.
The situation is the same if $\xi > 0$ and $1-\xi\phi^2(r) > 0$, but
we will have 
$\lim_{\bar{\phi}\rightarrow\infty} \phi = 1/\sqrt{\xi}$. If
$\xi\phi^2(r) -1> 0$ we can also apply the same formulation with minor
modifications. It is easy to verify that the transformation given by
\begin{eqnarray}
\label{trans3}
g_{\mu\nu} &=& (\xi\phi^2 -1)^{-\frac{2}{n-2}} \bar{g}_{\mu\nu}, \\
\label{trans4}
\bar{\phi}(\phi) &=& \int_a^\phi d\chi
\frac{\sqrt{1 + \xi\left(\frac{\xi}{\xi_c} - 1\right)\chi^2}}{\xi\chi^2-1},
\end{eqnarray}
maps also a solution $(g_{\mu\nu},\phi)$ of (\ref{action}) to a solution
$(\bar{g}_{\mu\nu},\bar{\phi})$ of (\ref{action2}).
However, we see from (\ref{trans4}) that in this case one needs also
that $\xi \ge \xi_c$. The integrals (\ref{trans2}) and (\ref{trans4}) 
can be explicitly solved, see \cite{AWKP} for instance, but the 
final expressions are rather cumbersome and in fact we will need only
some asymptotic expansions.
To summarize, we are able to generate all solutions
of (\ref{action}) for $\xi<0$. For $\xi >\xi_c$, we can generate all solutions
with $|\phi(r)| < 1/\sqrt{\xi}$ or with $|\phi(r)| > 1/\sqrt{\xi}$. Finally, 
for $0<\xi<\xi_c$ we can generate the solutions such that
$\frac{1}{\xi} < \phi^2(r) < \frac{\xi_c}{\xi(\xi_c-\xi)}$ 
or $|\phi(r)| < 1/\sqrt{\xi}$. Now we will examine 
each one of these special ranges of $\xi$.

For $\xi <0$, the transformations (\ref{trans1}) and (\ref{trans2}) can be
used for generating solutions with 
$\phi$ of any range. From (\ref{trans1}), we see that
the only candidate to null hyper-surface
for the metric ${g}_{\mu\nu}$ is that one for which $r=r_0$. To search
for black-hole solutions one needs 
to search for values of $\lambda$, $\xi$,
 and $n$ such that the hyper-surface
$r=r_0$ be a regular one of null type, and to this end 
we will evaluate the scalar of curvature $R$ (\ref{scal}) and $g_{tt}
$for $r=r_0$. 
Such a work
can be simplified considerably if one uses an asymptotic expansion for large 
$\phi$. From (\ref{trans2}) we have that  
$\bar{\phi} \approx {\sqrt{{1}/{\xi_c}-{1}/{\xi}}}\ln \phi$ 
for large $\phi$, and
from this we get that $\phi(r) \approx {\cal R}_n^{-\alpha}$
for $r = r_0 + \varepsilon$, where
\begin{equation} 
\label{alpha}
\alpha = \sqrt{\frac{\frac{n-2}{n-3}(1-\lambda^2)}
{{1}/{\xi_c}-{1}/{\xi}}}.
\end{equation}
Using that 
$g_{rr}\approx\left(1 + \frac{r_0^{n-3}}{r^{n-3}}\right)^{\frac{4}{n-3}}
{\cal R}_n^{2(\beta-\alpha)}$, where
\begin{equation}
\label{beta}
\beta = \frac{1-\lambda}{n-3}, 
\end{equation}
 and inserting the asymptotic expansion for $\phi$ in
(\ref{scal}) we get that, for $r = r_0 + \varepsilon$,
\begin{equation}
\label{scal1}
R \approx 
\frac{4\alpha^2(n-3)^2}{\xi^{\frac{n}{n-2}}}\times
\frac{r^{2n-4}r_0^{2n-6}}
{\left( r^{n-3}+r_0^{n-3}\right)^{\frac{4n-8}{n-3}}}\times 
{\cal R}_n^{-2\left(\frac{\alpha}{n-2} +\beta + 1\right)}.
\end{equation}
The expression (\ref{scal1}) has a non-removable singularity 
in $r=r_0$ for any $n>3$, $\xi<0$, and for any 
$\lambda \ne \pm 1$.
Thus, for such the hyper-surface $r=r_0$ is a naked-singularity and
this excludes the possibility that some solution does represent a black-hole. 
The only possibility of a black-hole
corresponds to the choice $\lambda=1$, as we see from the expression of
$g_{tt}$
\begin{equation}
\label{gtt1}
g_{tt} = - {\cal R}_n^{2\left( \frac{\alpha}{n-2} + \lambda\right)},
\end{equation}
what leads to $\phi=a$, and 
this solution is the usual $n$-dimensional Schwarzschild one.
For completeness, let us analyze the solution generated by the interior
solution of (\ref{solu}). Again, the
unique null hypersurface is $r=r_0$. We can check that the same
asymptotic expansions (\ref{scal1}) and (\ref{gtt1}) are valid for 
$r=r_0-\varepsilon$,  and from this we conclude also these solutions cannot 
give new black-holes.

We conclude that {\em
for $\xi<0$ and $n>3$ there is no other black-hole solution for
the action (\ref{action}) than the Schwarzschild one}.

For $\xi > \xi_c$ we are able to generate solutions with 
$|\phi(r)|<1/\sqrt{\xi}$ and with
$|\phi(r)|>1/\sqrt{\xi}$ by using the transformations
(\ref{trans1})-(\ref{trans2}) and (\ref{trans3})-(\ref{trans4})
respectively. We begin by the first case.
We can see from (\ref{trans1}) that the possible 
hyper-surface of null-type corresponds to that one for which
$r=r_0$. We also will examine the scalar of curvature to search for
horizons, but in this case we will use an asymptotic expansion for
(\ref{trans2}) valid for $\phi$ very close to $1/\sqrt{\xi}$.
For small $1-\xi\phi^2$ we obtain 
$\bar{\phi}(\phi)\approx -\sqrt{\frac{n-1}{n-2}} \ln 
\left( 
1-\xi\phi^2
\right) $, what leads to 
$1-\xi\phi^2 \approx {\cal R}_n^\delta$
for $r=r_0+\varepsilon$, where
\begin{equation}
\label{delta}
\delta=2\sqrt{\xi_c\frac{n-2}{n-3}(1-\lambda^2)}
\end{equation}
The derivative $\frac{d\phi}{dr}$ present in (\ref{scal}) can be 
evaluated for $r=r_0+\varepsilon$ by using  
$\frac{d\phi}{dr}=\frac{d\phi}{d\bar{\phi}}\frac{d\bar{\phi}}{dr}$
and calculating $\frac{d\phi}{d\bar{\phi}}$ from (\ref{trans2}) for
small $1-\xi\phi^2$. One gets
\begin{equation}
\frac{d\phi}{dr}\approx -2\sqrt{\frac{
\xi_c(n-3)(n-2)(1-\lambda^2)}{\xi}}\times 
\frac{r^{n-4}r_0^{n-3}}{(r^{n-3}+r_0^{n-3})^2}
\times{\cal R}_n^{\delta-1}.
\end{equation}
Using that $g_{rr}\approx\left(1 + 
\frac{r_0^{n-3}}{r^{n-3}}\right)^{\frac{4}{n-3}} {\cal R}_n^{
2\beta - \frac{2}{n-2}\delta}$ we finally get
\begin{equation}
\label{scal2}
R\approx 
\frac{4\xi_c(\xi_c -\xi)(n-3)(n-2)(1-\lambda^2)}{\xi^2}\times
\frac{r^{2n-4}r_0^{2n-6}}
{\left( r^{n-3}+r_0^{n-3}\right)^{\frac{4n-8}{n-3}}}\times
{\cal R}_n^{ 2\left(\frac{n-1}{n-2}\delta - \beta  - 1\right)}.
\end{equation}
Due to that $\frac{n-1}{n-2}\delta - \beta  - 1< 0$ we see that
(\ref{scal2}) is divergent for any $n>3$, $\xi>\xi_c$, 
and $\lambda \ne \pm 1$. For this case we have also
\begin{equation}
\label{gtt2}
g_{tt} = - {\cal R}_n^{2(\lambda - 2\delta)},
\end{equation}
discarding the possibility of $\lambda=-1$.
Again the only non-singular situation
is the usual vacuum solution.

The solutions with $|\phi(r)|>1/\sqrt{\xi}$ are generated by using the
transformations (\ref{trans3}) and (\ref{trans4}). The expressions for
$R$ and $g_{tt}$
valid for $r = r_0 \pm \varepsilon$ is also given, up to signs, 
by (\ref{scal1}) and (\ref{gtt1}),
what excludes any black-hole solution for $|\phi|>1/\sqrt{\xi}$ and
$\lambda\ne 1$.

We have that {\em
for $\xi>\xi_c$ and $n>3$ there is no other black-hole solution 
with the scalar
field obeying $|\phi(r)|<1/\sqrt{\xi}$ or $|\phi(r)|>1/\sqrt{\xi}$
than the Schwarzschild one.}

Finally we have the case $0<\xi<\xi_c$. Solutions for which
$|\phi(r)|<1/\sqrt{\xi}$ are generated by (\ref{trans1}) and (\ref{trans2}),
and
the asymptotic expressions for $R$ and $g_{tt}$ are (\ref{scal2})
and (\ref{gtt2}) respectively. For the range
$\frac{1}{\xi} < \phi^2(r) < \frac{\xi_c}{\xi(\xi_c-\xi)}$ the asymptotic
expressions for $R$ and $g_{tt}$ are also given by (\ref{scal1})
and (\ref{gtt1}). Thus, we conclude
again that {\em
for $0<\xi<\xi_c$ and $n>3$ there is no other black-hole solution 
 with the scalar
field obeying $|\phi(r)|<1/\sqrt{\xi}$ or 
$\frac{1}{\xi} < \phi^2(r) < \frac{\xi_c}{\xi(\xi_c-\xi)}$ 
than the Schwarzschild one.}

We finish saying that our ``no-go'' results for scalar hairs with
arbitrary coupling buttresses the recent conclusions that the
the four dimensional Bekenstein black-hole solution is truly
exclusive and outstanding\cite{Z,XZ,xd,CK}. If we remember that
the Bekenstein solution has the same number of free parameters
as the Reissner-Nordstr\"om solution and not one more\cite{XZ},
we can say that the essence of the no-hair conjecture is not
compromised by the conformal scalar hair.

\end{document}